\RequirePackage{fix-cm}
\documentclass[english,notitlepage,prl,reprint,unsortedaddress]{revtex4-1}
\usepackage[latin9]{inputenc}
\usepackage{amsmath}
\usepackage{graphicx}

\makeatletter

\DeclareFontEncoding{LGR}{}{}
\DeclareRobustCommand{\greektext}{%
  \fontencoding{LGR}\selectfont\def\encodingdefault{LGR}}
\DeclareRobustCommand{\textgreek}[1]{\leavevmode{\greektext #1}}
\ProvideTextCommand{\~}{LGR}[1]{\char126#1}

\makeatother

\usepackage{babel}
\begin{document}
\title{Stochastic density functional theory }
\author{Marcel David Fabian}
\email{These authors contributed equally}

\affiliation{Fritz Haber Center for Molecular Dynamics and Institute of Chemistry,
The Hebrew University of Jerusalem, Jerusalem 9190401, Israel}
\author{Ben Shpiro}
\email{These authors contributed equally}

\affiliation{Fritz Haber Center for Molecular Dynamics and Institute of Chemistry,
The Hebrew University of Jerusalem, Jerusalem 9190401, Israel}
\author{Eran Rabani}
\email{eran.rabani@berkeley.edu}

\affiliation{Department of Chemistry, University of California, Berkeley, California
94720, USA, and Materials Science Division, Lawrence Berkeley National
Laboratory, Berkeley, California 94720, USA, and The Sackler Center
for Computational Molecular and Materials Science, Tel Aviv University,
Tel Aviv 69978, Israel}
\author{Daniel Neuhauser}
\email{dxn@ucla.edu}

\affiliation{Department of Chemistry and Biochemistry, University of California,
Los Angeles California 90095, USA}
\author{Roi Baer}
\email{roi.baer@huji.ac.il}

\affiliation{Fritz Haber Center for Molecular Dynamics and Institute of Chemistry,
The Hebrew University of Jerusalem, Jerusalem 9190401, Israel}
\begin{abstract}
Linear-scaling implementations of density functional theory (DFT)
reach their intended efficiency regime only when applied to systems
having a physical size larger than the range of their Kohn-Sham density
matrix (DM). This causes a problem since many types of large systems
of interest have a rather broad DM range and are therefore not amenable
to analysis using DFT methods. For this reason, the recently proposed
stochastic DFT (sDFT), avoiding exhaustive DM evaluations, is emerging
as an attractive alternative linear-scaling approach. This review
develops a general formulation of sDFT in terms of a (non)orthogonal
basis representation and offers an analysis of the statistical errors
(SEs) involved in the calculation. Using a new Gaussian-type basis-set
implementation of sDFT, applied to water clusters and silicon nanocrystals,
it demonstrates and explains how the standard deviation and the bias
depend on the sampling rate and the system size in various types of
calculations. We also develop basis-set embedded-fragments theory,
demonstrating its utility for reducing the SEs for energy, density
of states and nuclear force calculations. Finally, we discuss the
algorithmic complexity of sDFT, showing it has CPU wall-time linear-scaling.
The method parallelizes well over distributed processors with good
scalability and therefore may find use in the upcoming exascale computing
architectures.
\end{abstract}
\maketitle

\section{Introduction}

Density functional theory (DFT) is emerging as a usefully-accurate
general-purpose computational platform for predicting from first principles
the ground-state structure and properties of systems spanning a wide
range of length scales, from single atoms and gas-phase molecules,
through macromolecules, proteins, nanocrystals, nanosheets, nanoribbons,
surfaces, interfaces up to periodic or amorphous homogeneous or heterogeneous
materials \citep{Ratcliff2013,tsuneda2014density,morin2013density,engel2011density,Graziani2014}.
Significant efforts have been diverted towards the development of
numerical and computational methods enabling the use of DFT for studying
extensive molecular systems. Several routes have been suggested: linear-scaling
approaches \citep{Yang1991,Li1993,Ordejon1993,Goedecker1994,Nunes1994,Wang1995c,Hernandez1995,goedecker1995lowcomplexity,Ordejon1996,Bowler1997,Baer1997b,Baer1997a,Palser1998,Goedecker1999,Scuseria1999,Galli2000,Adhikari2001,Soler2002,Skylaris2005a,Gillan2007,Ochsenfeld2007,Havu2009,lin2009polebased,Ozaki2010,bowler2012methods,moussa2016minimax,Ratcliff2017},
relying on the sparsity of the density matrix \citep{Kohn1996}, DFT-based
tight-binding (DFTB) methods \citep{Elstner1998,Aradi2007,karasiev2015frank}
which reduce the numerical scaling using model Hamiltonians. Moreover,
significant efforts have gone towards developing orbital-free DFT
\citep{Witt2018,karasiev2015frank} approaches using density-dependent
kinetic energy functionals. The first two types of approaches mentioned
above are designed to answer questions typically asked about molecules,
while for materials and other large scale systems, we are more interested
in coarse-grained properties. For example, with molecules, one is
interested in bond orders, bond lengths spectral lines; while for
large systems we are more interested in atomic densities, pair-correlation
distributions (measured using neutron scattering) as well as charge/spin
densities, polarizabilities and optical and electrical conductivity.
In molecules, we strive to understand each occupied/unoccupied Kohn-Sham
eigenstate while in large systems we are concerned with the density
of hole and electron states. 

Of course, detailed ``molecular type'' questions can also arise
in large systems, primarily when the processes of interest occur in
small pockets or localized regions --- for example, biochemical processes
in proteins, localized catalytic events on a surface, impurities in
solids, etc. Here, a combination of methods, where the small subsystem
can be embedded in the larger environment is required. 

In this advanced review, we will focus on the stochastic DFT (sDFT)
approach, developed using grids and plane-waves in recent years \citep{Baer2013,Neuhauser2014a,arnon2017equilibrium,Cytter2018,Neuhauser2015}
but also based on ideas taken from works starting in the early 1990's,
mainly within the tight-binding electronic structure framework \citep{Drabold1993,Sankey1994,Wang1994g,Roeder1997,weisse2006kernel,bekas2007estimator,lin2016approximating,wang2018gradient}.
We make the point that the efficiency of sDFT results from its adherence
to answering the coarse-grained \textquotedblleft large system questions\textquotedblright{}
mentioned above, rather than those asked for molecules. 

The new viewpoint taken here is that of stochastic DFT using non-orthogonal
localized basis-sets. The primary motivation behind choosing local
basis-sets is that they are considerably more compact than plane-waves
and therefore may enable studying significantly larger systems. Deterministic
calculations using local basis-sets are more readily applicable to
large systems, and thus can generate useful benchmarks with which
the statistical errors and other properties characterizing sDFT can
be studied in detail. 

The review includes three additional sections, further divided into
subsections, to be described later. Section~\ref{sec:Theory-and-methods}
reviews the theory and techniques used for non-orthogonal sDFT and
studies in detail the statistical errors and their dependence on sampling
and system size. In section~\ref{sec:Embedded-fragments-based} we
explain the use of embedded fragments and show their efficacy in reducing
the stochastic errors of sDFT. Section~\ref{sec:Summary} summarizes
and discusses the findings. 

\section{\label{sec:Theory-and-methods}Theory and methods}

In this section, we discuss three formulations of KS-DFT represented
in non-orthogonal basis-sets. Since the issue of algorithmic scaling
is at the heart of developing DFT methods for large systems, we emphasize
for each formulation the associated algorithmic complexity (so-called
system-size scaling). We start with the traditional basis-set formulation
of the Kohn-Sham equations leading to standard cubic-scaling (subsection~\ref{subsec:cubicDFT}).
Then, showing how, by focusing on observables and exploiting the sparsity
of the matrices, a quadratic-scaling approach can be developed with
no essential loss of rigor or accuracy (subsection~\ref{subsec:quadraticDFT}).
Most of the discussion will revolve around the third and final approach,
stochastic DFT, which estimates expectation values using stochastic
sampling methods, as described in subsection~\ref{subsec:linearDFT}.
This latter approach leads, to linear-scaling complexity. 

\subsection{\label{subsec:cubicDFT}Traditional basis-set formulation of Kohn-Sham
equations with cubic scaling}

The Kohn-Sham (KS) density functional theory (KS-DFT) is a molecular
orbitals (MOs) approach which can be applied to a molecular system
of $N_{e}$ electrons using a basis-set of atom-centered orbitals
$\phi_{\alpha}\left(\boldsymbol{r}\right)$, $\alpha=1,\dots,K.$
The basis functions were developed to describe the electronic structure
of the parent atom, and for molecules they are the building blocks
from which the orthonormal MOs are built as superpositions:

\begin{equation}
\psi_{n}\left(\boldsymbol{r}\right)=\sum_{\alpha=1}^{K}\phi_{\alpha}\left(\boldsymbol{r}\right)C_{\alpha n},\,n=1,\dots,K.\label{eq:MO-def}
\end{equation}
In the simplest ``population'' model, each MO can either ``occupy''
two electrons (of opposing spin) or be empty. The occupied MOs (indexed
as the first $N_{occ}=N_{e}/2$ MOs) are used to form the total electron
density:
\begin{equation}
n\left(\boldsymbol{r}\right)=2\times\sum_{n}^{N_{occ}}\left|\psi_{n}\left(\boldsymbol{r}\right)\right|^{2}.\label{eq:ChargeDensity}
\end{equation}

The coefficient matrix $C$ in Eq.~(\ref{eq:MO-def}) can be obtained
from the variational principle applied to the Schr�dinger equation,
leading to the Roothaan-Hall generalized eigenvalue equations \citep{Roothaan1951,Hall1951}
(we follow the notations in refs.~\citep{Szabo1996,Koch2001,Abrol2002}):
\begin{equation}
FC=SCE.\label{eq:Roothan}
\end{equation}
Here, $F=T+V^{en}+J\left[n\right]+V^{xc}\left[n\right]$ is the $K\times K$
KS Fock matrix, $S_{\alpha\alpha'}=\left\langle \phi_{\alpha}|\phi_{\alpha'}\right\rangle $
is the overlap matrix of the AO's and $E$ is a diagonal matrix containing
the MO energies, $\varepsilon_{1},\dots,\varepsilon_{K}$. The Fock
matrix $F_{\alpha\alpha'}$ includes the kinetic energy integrals,
$T_{\alpha\alpha'}=\left\langle \phi_{\alpha}\left|-\frac{1}{2}\nabla^{2}\right|\phi_{\alpha'}\right\rangle $,
the nuclear attraction integrals $V_{\alpha\alpha'}^{en}=\left\langle \phi_{\alpha}\left|\hat{v}_{en}\right|\phi_{\alpha'}\right\rangle $,
where $\hat{v}_{en}$ is the electron-nuclear interaction operator,
the Coulomb integrals $J_{\alpha\alpha'}=\left\langle \phi_{\alpha}\left|v_{H}\left[n\right]\left(\hat{\boldsymbol{r}}\right)\right|\phi_{\alpha'}\right\rangle $,
where $v_{H}\left[n\right]\left(\boldsymbol{r}\right)=\int\frac{n\left(\boldsymbol{r}'\right)}{\left|\boldsymbol{r}-\boldsymbol{r}'\right|}d^{3}r'$
is the Hartree potential, and finally, the exchange-correlation integrals,
$V_{\alpha\alpha'}^{xc}=\left\langle \phi_{\alpha}\left|v_{xc}\left[n\right]\left(\hat{\boldsymbol{r}}\right)\right|\phi_{\alpha'}\right\rangle $
where $v_{xc}\left[n\right]\left(\hat{\boldsymbol{r}}\right)$ is
the exchange correlation potential.

In KS theory, the Fock matrix $F$ and the electron density $n\left(\boldsymbol{r}\right)$
are mutually dependent on each other and must be obtained self-consistently.
This is usually achieved by converging an iterative procedure, 
\begin{align}
\dots\longrightarrow n\left(\boldsymbol{r}\right) & \longrightarrow\left\{ v_{H}\left[n\right]\left(\boldsymbol{r}\right),v_{xc}\left[n\right]\left(\boldsymbol{r}\right)\right\} \longrightarrow\label{eq:SCF-O(K^3)}\\
\longrightarrow F & \xrightarrow{O\left(K^{3}\right)}\left\{ C,E\right\} \longrightarrow n\left(\boldsymbol{r}\right)\longrightarrow\dots,\nonumber 
\end{align}
where in each iteration, a previous density iterate $n\left(\boldsymbol{r}\right)$
is used to generate the Hartree $v_{H}\left[n\right]\left(\boldsymbol{r}\right)$
and exchange-correlation $v_{xc}\left[n\right]\left(\boldsymbol{r}\right)$
potentials from which we construct the Fock matrix $F$. Then, by
solving Eq.~(\ref{eq:Roothan}) the coefficient matrix $C$ is obtained
from which a new density iterate $n\left(\boldsymbol{r}\right)$ is
generated via Eqs.~(\ref{eq:MO-def})-(\ref{eq:ChargeDensity}).
The iterations continue until convergence (density stops changing
with a predetermined threshold), and a \emph{self-consistent field
solution }is thus obtained. 

This implementation of the basis-set based approach becomes computationally
expensive for very large systems due to the cubic scaling of solving
the algebraic Roothan-Hall equations (Eq.~(\ref{eq:Roothan})). This
cubic-scaling step is marked by placing $O\left(K^{3}\right)$ on
the corresponding arrow in Eq.~(\ref{eq:SCF-O(K^3)}). The Coulomb
integral calculation has a much lower scaling and can be completed
in a $O\left(K\log K\right)$ scaling effort, either using continuous
fast-multipole methods \citep{Shao2000,white1994continuous} or fast-Fourier
transforms on grids, as done here.

\subsection{\label{subsec:quadraticDFT}Equivalent trace-based formulation with
quadratic scaling}

In order to lower the scaling, we can take advantage of the fact that
both $F$ and $S$ are very sparse matrices in the AO representation.
The complication, however, is that the $C$ matrix of Eq.~(\ref{eq:Roothan})
is non-sparse and therefore should be circumvented. This is challenging
since the $C$ matrix of Eq.~(\ref{eq:Roothan}) is used to extract
both the eigenvalues $\varepsilon_{n}$ and at the same time to enforce
the MO orthogonalization, both described by the matrix equations:
\begin{equation}
C^{T}FC=E,\,\,\text{and}\,\,C^{T}SC=I.\label{eq:CTFC}
\end{equation}

The first step in circumventing the calculation of the $C$ matrix
introduces the \emph{density matrix }(DM) formally defined as 
\begin{equation}
P=Cf\left(E;T,\mu\right)C^{T},\label{eq:P=00003DCfC^T}
\end{equation}
where $f\left(E;T,\mu\right)$ is the diagonal matrix obtained by
plugging $E$ instead of $\varepsilon$ in the Fermi-Dirac distribution
function: 
\begin{equation}
f\left(\varepsilon;T,\mu\right)\equiv\frac{1}{1+e^{\left(\varepsilon-\mu\right)/k_{B}T}}.\label{eq:FD}
\end{equation}
The diagonal matrix elements, $2f\left(\varepsilon_{n}\right)$ (we
omit designating the temperature $T$ and chemical $\mu$ in $f$
when no confusion is expected) represent the level occupation of the
MO $\psi_{n}\left(\boldsymbol{r}\right)$ (which typically holds a
spin-up and a spin-down electron, hence the factor of 2). $T$ can
be a real finite temperature or a very low fictitious one. In the
latter case, the $T\to0$ limit of Eq.~(\ref{eq:FD}) yields $f\left(\varepsilon_{n}\right)=1$
for $n\le N_{\text{occ}}$ and $0$ otherwise, assuming that the chemical
$\mu$ has been chosen such that $N_{e}=2\sum_{n}f\left(\varepsilon_{n}\right)$. 

In contrast to the formal definition in Eq.~(\ref{eq:CTFC}) of $P$
as a matrix, in sDFT regards $P$ as an \emph{operator }expressed
in terms of $F$ and $S$ through the relation 
\begin{equation}
P=f\left(S^{-1}F;T,\mu\right)S^{-1}.\label{eq:P=00003Df(S^-1F)S^-1}
\end{equation}
Here, $S^{-1}F$ is ``plugged'' in place of $\varepsilon$ into
the function $f$ of Eq.~(\ref{eq:FD}) \footnote{This relation can be proved by plugging $E=C^{T}FC$ from Eq.~(\ref{eq:CTFC})
into Eq.~(\ref{eq:P=00003DCfC^T}), giving $P=Cf\left(C^{T}FC;T,\mu\right)C^{T}$,
then using the rule $Af\left(XA\right)=f\left(AX\right)A$ (valid
for functions that can be represented as power series and square matrices)
obtain $P=f\left(CC^{T}F;T,\mu\right)CC^{T}$ and finally using $CC^{T}=S^{-1}$
from from Eq.~(\ref{eq:CTFC}).}. Just like $P$ is an operator, our method also views $S^{-1}$ as
an \emph{operator }which is applied to any vector $u$ with linear-scaling
cost using a preconditioned conjugate gradient method {[}59,60{]}).
The \emph{operator }$P$, applied to an arbitrary vector $u$, uses
a Chebyshev expansion {[}9,17,44,61{]} of length $N_{C}$: $Pu=\sum_{l=0}^{N_{C}}a_{l}\left(T,\mu\right)u^{l}$
where $a_{l}$ are the expansion coefficients and $u^{0}=S^{-1}u$,
$u^{1}=Hu^{0}$ and then $u^{l+1}=2Hu^{l}-u^{l-1}$, $l=2,3,...$.
In this expansion the operator $H$ is a shifted-scaled version of
the operator $S^{-1}F$ bringing its eigenvalue spectrum into the
$\left[-1,1\right]$ interval. Every operation $Pu$, which involves
repeated applications of $H$ to various vectors is automatically
linear-scaling due to the fact that $F$ and $S$ are sparse. Clearly,
the numerical effort in the application of $P$ to $u$ depends on
the length $N_{C}$ of the expansion. When the calculation involves
a finite physical temperature $T$, $N_{C}=2\left(\frac{E_{max}-E_{min}}{k_{B}T}\right)$,
where $E_{max}\left(E_{min}\right)$ is the largest (smallest) eigenvalue
of $H$. Since $N_{C}$ is inversely proportional to $T$, the numerical
effort of sDFT reduces as $T^{-1}$ in contrast to deterministic KS-DFT
approaches where it rises as $T^{3}$ \citep{Cytter2018}. For zero
temperature calculations one still uses a finite temperature but chooses
it according to the criterion $k_{B}T\ll\varepsilon_{g}$ where $\varepsilon_{g}$
is the KS energy gap. For metals it is common to take a fictitious
low temperatures. 

The above analysis shows then, that the application of $P$ to a vector
can be performed in a linear-scaling cost without constructing $P$.
We use this insight in combination with the fact that the expectation
value of one-body observables $\hat{O}=\sum_{n=1}^{N_{e}}\hat{o}_{n}$
(where $\hat{o}$ is the underlying single electron operator and the
sum is over all electrons) can be achieved as a matrix trace with
$P$:
\begin{equation}
\left\langle \hat{O}\right\rangle =2\text{Tr}\left[PO\right],\label{eq:<O>=00003Dtr=00005BPO=00005D}
\end{equation}
where $O_{\alpha\alpha'}=\left\langle \phi_{\alpha}\left|\hat{o}\right|\phi_{\alpha'}\right\rangle $
is the matrix representation of the operator within the atomic basis.
Eq.~(\ref{eq:<O>=00003Dtr=00005BPO=00005D}) can be used to express
various expectation values, such as the electron number
\begin{align}
N_{e} & =2\text{Tr}\left[PS\right]\label{eq: Ne-trace}\\
 & =2Tr\left[f\left(S^{-1}F;T,\mu\right)\right],\nonumber 
\end{align}
the orbital energy 
\begin{align}
E_{orb} & =2Tr\left[PF\right]\label{eq:Eorb-trace}\\
 & =2\text{Tr}\left[e\left(S^{-1}F;T,\mu\right)\right],\nonumber 
\end{align}
where, $e\left(\varepsilon\right)=f\left(\varepsilon\right)\varepsilon$
and the fermionic entropy 
\begin{align}
\Sigma & _{F}=-2k_{B}\text{Tr}\left[PS\ln PS+\left(I-PS\right)\ln\left(I-PS\right)\right]\label{eq:S-trace}\\
 & =2\text{Tr}\left[\sigma_{F}\left(S^{-1}F;\beta,\mu\right)\right]\nonumber 
\end{align}
where $\sigma_{F}=-k_{B}\left(f\ln f+\left(1-f\right)\ln\left(1-f\right)\right)$.
The expectation value of another observable, the density of states
$\rho_{s}\left(E\right)=\sum_{n}\delta\left(E-\varepsilon_{n}\right)$
can also be written as a trace \citep{Gross1991}:
\begin{align}
\rho_{s}\left(\varepsilon\right) & =\pi^{-1}\lim_{\eta\to0}\text{Im}\,\text{Tr}\left[\left(\varepsilon S-F-i\eta S\right)^{-1}S\right],\label{eq:DOS-trace}\\
 & =\pi^{-1}\text{Im}\,\text{Tr}\left[g\left(S^{-1}F;\varepsilon\right)\right]\nonumber 
\end{align}
where $g\left(\varepsilon';\varepsilon\right)=\lim_{\eta\to0}\frac{1}{\varepsilon-\varepsilon'-i\eta}$.

Since the density matrix is an operator in the present approach, the
trace in Eq.~(\ref{eq:<O>=00003Dtr=00005BPO=00005D}) can be evaluated
by introducing the unit column vectors $u^{\left(\alpha'\right)}$
($\alpha'=1,\dots,K$) and operating with $P$ on them, and the trace
becomes:
\begin{equation}
\left\langle \hat{O}\right\rangle =2\sum_{\alpha,\alpha'=1}^{K}\left(Pu^{\left(\alpha'\right)}\right)_{\alpha}O_{\alpha\alpha'}.\label{eq:O-from-Pu}
\end{equation}
Evaluating this equation requires \emph{quadratic-scaling} computational
complexity since it involves $K$ applications of $P$ to unit vectors
$u^{\left(\alpha'\right)}$ . One important use of Eq.~(\ref{eq:<O>=00003Dtr=00005BPO=00005D})
is to compute the electron density at spatial point $\boldsymbol{r}$: 

\begin{align}
n\left(\boldsymbol{r}\right) & =2\text{Tr}\left[PN\left(\boldsymbol{r}\right)\right],\label{eq:DensityAsATrace}
\end{align}
where $N_{\alpha\alpha'}\left(\boldsymbol{r}\right)=\phi_{\alpha}\left(\boldsymbol{r}\right)\phi_{\alpha'}\left(\boldsymbol{r}\right)$
is the overlap distribution matrix, leading to the expression 
\begin{equation}
n\left(\boldsymbol{r}\right)=2\sum_{\alpha,\alpha'=1}^{K}\left(Pu^{\left(\alpha'\right)}\right)_{\alpha}\phi_{\alpha}\left(\boldsymbol{r}\right)\phi_{\alpha'}\left(\boldsymbol{r}\right).\label{eq:densityFromP}
\end{equation}
Here, given $\boldsymbol{r}$, only a finite (system-size independent)
number of $\alpha$ and $\alpha'$ pairs must be summed over. Hence,
the calculation of the density at just this point involves a linear-scaling
effort because of the need to apply $P$ to a finite number of $u^{\left(\alpha'\right)}$'s.
It follows, that the density function $n\left(\boldsymbol{r}\right)$
on the entire grid can be obtained in quadratic scaling effort \footnote{Note that when the DM $P$ is sparse, the evaluation of the density
of Eq.$\ $\ref{eq:densityFromP} can be performed in linear-scaling
complexity. The stochastic method (explained in Subsection~\ref{subsec:sDFT-Formulation})
does not exploit this sparsity explicitly.}. This allows us to change the SCF schema of Eq.~(\ref{eq:SCF-O(K^3)})
to:
\begin{align}
\dots\longrightarrow n\left(\boldsymbol{r}\right) & \longrightarrow\left\{ v_{H}\left[n\right]\left(\boldsymbol{r}\right),v_{xc}\left[n\right]\left(\boldsymbol{r}\right)\right\} \longrightarrow\label{eq:SCF-O(K^2)}\\
\longrightarrow F & \xrightarrow{O\left(K^{2}\right)}n\left(\boldsymbol{r}\right)\longrightarrow\dots,\nonumber 
\end{align}
where the quadratic step is marked $O\left(K^{2}\right)$. 

Summarizing, we have shown an alternative trace-based formulation
of Kohn Sham theory which focuses on the ability to apply the DM to
vectors in a linear-scaling way, without actually calculating the
matrix $P$ itself. This leads to a deterministic implementation of
KS-DFT theory of quadratic scaling complexity.

\subsection{\label{subsec:linearDFT}Basis-set stochastic density functional
theory with linear-scaling}

The first report of linear-scaling stochastic DFT (sDFT) \citep{Baer2013}
used a grid-based implementation and focused on the standard deviation
error. Other developments of sDFT included implementation of a stochastic
approach to exact exchange in range-separated hybrid functionals \citep{Neuhauser2015}
and periodic plane-waves applications to warm dense matter \citep{Cytter2018}
and materials science \citep{Ming2018}. These developments were all
done using orthogonal or grid representations and included limited
discussions of the statistical errors. 

Here, sDFT is presented in a general way (subsection~\ref{subsec:sDFT-Formulation}),
applicable to any basis, orthogonal or not. We then present a theoretical
investigation of the variance (subsection~\ref{subsec:fluctuations})
and bias (subsection~\ref{subsec:bias}) errors, and using our Gaussian-type
basis code, bsInbar, we actually calculate these SEs in water clusters
\footnote{The clusters we used were produced by Daniel Sp�ngberg at Uppsala
University, Department of Materials Chemistry, and retrieved from
the ergoscf webpage http://www.ergoscf.org/xyz/h2o.php.} (by direct comparison to the deterministic results) and study their
behavior with sampling and system size. Finally, in subsection~\ref{subsec:Scaling-and-scalability}
we discuss the scaling and the scalability of the method.

\subsubsection{\label{subsec:sDFT-Formulation}sDFT formulation}

Having described the quadratic scaling in the previous section, we
are but a step away from understanding the way sDFT works. The basic
idea is to evaluate the trace expressions (Eqs.~(\ref{eq:<O>=00003Dtr=00005BPO=00005D})-(\ref{eq:densityFromP}))
using the \emph{stochastic trace formula }\citep{Hutchinson1990}:
\begin{equation}
\text{Tr}\left[M\right]=\text{\textbf{E} }\left\{ \sum_{\alpha\alpha'}^{K}\chi_{\alpha}M_{\alpha\alpha'}\chi_{\alpha'}\right\} \equiv\text{\textbf{E} }\left\{ \chi^{T}M\chi\right\} ,\label{eq:HutchinsonTrace}
\end{equation}
where $M$ is an arbitrary matrix, $\chi_{\alpha}$ are $K$ random
variables taking the values $\pm1$ and $\text{\textbf{E} }\left\{ \chi^{T}M\chi\right\} $
symbolizes the statistical expected value of the functional $\chi^{T}M\chi$.
One should notice that Eq.~(\ref{eq:HutchinsonTrace}) is an identity,
since we actually take the expected value. However, in practice we
must take a finite sample of only $I$ independent random vectors
$\chi$'s. This gives an \emph{approximate} practical way of calculating
the trace of $M$:

\begin{equation}
\text{Tr}\left[M\right]\approx\text{Tr}_{I}\left[M\right]\equiv\frac{1}{I}\sum_{i=1}^{I}\left(\chi^{i}\right)^{T}M\chi^{i}.\label{eq:Trace-Sample}
\end{equation}
From the central limit theorem, this trace evaluation introduces a
fluctuation error equal to 
\begin{equation}
\boldsymbol{\text{Var}}\left(\text{Tr}_{I}\left[M\right]\right)=\frac{\Sigma_{M}^{2}}{I},
\end{equation}
where $\Sigma_{M}^{2}=\text{\textbf{Var}}\left(\text{Tr}_{1}\left[M\right]\right)$
is the variance of $\sum_{\alpha\alpha'}^{K}\chi_{\alpha}M_{\alpha\alpha'}\chi_{\alpha'}$
(discussed in detail in below). This allows to balance between statistical
fluctuations and numerical effort, a trade-off which we exploit in
sDFT. 

With this stochastic technique, the expectation value of an operator
$\hat{O}$ becomes (c.f. Eq.~(\ref{eq:O-from-Pu})):
\begin{equation}
\left\langle \hat{O}\right\rangle =2\text{\textbf{E} }\left\{ \left(P\chi\right)^{T}\left(O\chi\right)\right\} ,\label{eq:<O>=00003D2E=00007Bchi^T*PO*chi=00007D}
\end{equation}
where the application of $P$ to the random vector $\chi$ is performed
in the same manner as described above for $u$ (see the text immediately
after Eq.~(\ref{eq:P=00003Df(S^-1F)S^-1})). This gives the electronic
density (see Eq.~(\ref{eq:densityFromP})):

\begin{align}
n\left(\boldsymbol{r}\right) & =2\text{\textbf{E}}\left\{ \psi_{P\chi}\left(\boldsymbol{r}\right)\psi_{\chi}\left(\boldsymbol{r}\right)\right\} ,\label{eq:DensityAsStochTrace}
\end{align}
yielding a vector (called a grid-vector) of density values $n\left(\boldsymbol{r}\right)$
at each grid-point. This involves producing two grid-vectors, $\psi_{\chi}\left(\boldsymbol{r}\right)=\chi_{\alpha}\phi_{\alpha}\left(\boldsymbol{r}\right)$
and $\psi_{P\chi}\left(\boldsymbol{r}\right)=\left(P\chi\right)_{\alpha'}\phi_{\alpha'}\left(\boldsymbol{r}\right)$
and then multiplying them point by point and averaging on the $I$
random vectors.

\subsubsection{\label{subsec:-bsDFT-detail}sDFT calculation detail in the basis-set
formalism}

It is perhaps worthwhile discussing one trick-of-the-trade allowing
the efficient calculation of expectation values of some observables,
such as $N_{e}$, $E_{orb}$, $\Sigma_{F}$ and $\rho_{s}$, see Eqs.~(\ref{eq: Ne-trace})
- (\ref{eq:DOS-trace}). These are all expressed as traces over a
function $z\left(\varepsilon\right)$, respectively $f$$\left(\varepsilon\right)$,
$\varepsilon f\left(\varepsilon\right)$, $\sigma_{F}\left(\varepsilon\right)$
and $\rho_{e}\left(\varepsilon\right)$. As a result, all calculations
of such expectation values can be expressed as 
\begin{equation}
\text{Tr}\left[z\left(FS^{-1}\right)\right]=\sum_{l=0}^{N_{C}}a_{l}m_{l},\label{eq:trace=00005Bz(hN)=00005D}
\end{equation}
where $a_{l}$ are the Chebyshev expansion coefficients (defined above,
in subsection~\ref{subsec:quadraticDFT}), easily calculable, depending
on the function $z$ and:
\begin{equation}
m_{l}=\text{Tr}\left[\chi^{T}T_{l}\chi\right]=\text{\textbf{E}}\left\{ \chi^{T}T_{l}\chi\right\} ,\label{eq:StochasticMoments}
\end{equation}
are the Chebyshev moments \citep{Roeder1997}, where $T_{l}$ is the
$l$'th Chebyshev polynomial. The computationally expensive part of
the calculation, evaluating the moments $m_{l}$, is done once and
then used repeatedly for all relevant expectation values. One frequent
use of this moments method involves repeated evaluation of the number
of electrons $N_{e}$ until the proper value of the chemical potential
is determined.

We should note that many types of expectation values cannot be calculated
directly from the moments $m_{l}$. For example, the density, the
kinetic and potential energies. For these a full stochastic evaluation
is needed.

\begin{figure}
\begin{centering}
\includegraphics[width=1\columnwidth]{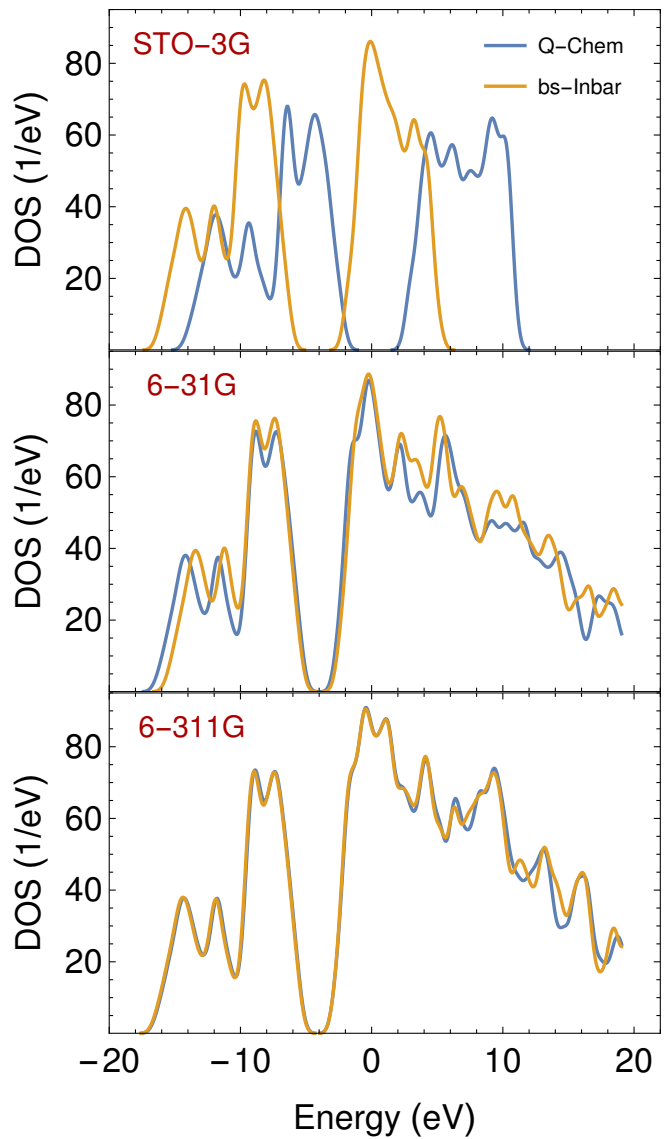}
\par\end{centering}
\caption{\label{fig:bsInbarValidation} The DOS as a function of energy for
a hydrogen-saturated silicon cluster ($\text{Si}_{87}\text{H}_{76}$)
calculated using the all-electron Q-CHEM \citep{Shao2015} and the
the bs-Inbar codes. Comparison is made for three standard Gaussian
basis-sets as indicated in the panels. We used the local density approximation
(LDA) for the exchange-correlation energy. Both calculations plot
the DOS of Eq.~(\ref{eq:DOS-trace}) using $k_{B}T=0.01E_{h}$. }
\end{figure}
Let us digress a little to explain how we make the calculations, presented
in this review, that enable us to study the properties of sDFT and
compare them to deterministic calculations. The code we have written
for that purpose is called \emph{bs-Inbar} \footnote{bs-Inbar is the basis-set version of our electronic structure program
named ``Inbar''. \emph{Inbar }is the Hebrew equivalent of the Greek
\textgreek{>'hlektron}, i.e. electron, which like Ambar and Amber,
of Perisan/Arabic origins, refers to the yellowish glowing fossilized
tree resin.}, and implements both the deterministic Kohn-Sham DFT approach described
in the present and previous sections as well as the stochastic DFT
to be discussed below. Following previous works \citep{Soler2002,VandeVondele2005},
we use an auxiliary equally-spaced (grid spacing $\Delta x=0.5a_{0}$)
Cartesian grid for calculating the electron-nuclear interaction integrals
$V_{\alpha\alpha'}^{en}$, the Coulomb repulsion integrals $J_{\alpha\alpha'}$,
built from the grid vector representing the density $n\left(\boldsymbol{r}\right)$
using fast Fourier transform techniques, and the exchange correlation
integrals $V_{\alpha\alpha'}^{xc}$. This is the $n\left(\boldsymbol{r}\right)\to\left\{ v_{H}\left[n\right]\left(\boldsymbol{r}\right),v_{xc}\left[n\right]\left(\boldsymbol{r}\right)\right\} \to F$
step of Eq.~(\ref{eq:SCF-O(K^3)}). We developed efficient methods
to represent the basis functions on the grid to quickly generate molecular
orbitals of the type of Eq.~(\ref{eq:MO-def}) on the grid. These
techniques are necessary for the step $F\to n\left(\boldsymbol{r}\right)$
of Eq.~(\ref{eq:SCF-O(K^2)}) for generating the density $n\boldsymbol{\left(r\right)}$
from the DM Eq.~(\ref{eq:densityFromP}). There are some technical
details, such as the effects of core electrons, which cannot be treated
efficiently on the grid, and thus are taken into account using norm-conserving
pseudopotentials techniques \citep{Troullier1991,Kleinman1982}, and
the deleterious Coulomb/Ewald images which are screened out using
the method of Ref.~\onlinecite{Martyna1999}. Additional technical
elements concerning the bs-Inbar implementation will be presented
elsewhere. In Fig.~\ref{fig:bsInbarValidation} we demonstrate the
validity of the deterministic bs-Inbar implementation by comparing
its $\text{Si}_{87}\text{H}_{76}$ DOS function to that obtained from
the eigenvalues of an all-electron calculation within the same basis-set
(using the Q-CHEM program \citep{Shao2015}). For the largest basis-set
(triple zeta 6-311G) the two codes produce almost identical DOS (with
small difference at high energies), while for the smallest basis (STO-3G)
the all electron result shifts strongly to higher energies. Clearly,
the bs-Inbar results are less sensitive to the basis-set, likely due
to the use of pseudopotentials instead of treating core electrons
explicitly. 

\begin{figure}
\begin{centering}
\includegraphics[width=1\columnwidth]{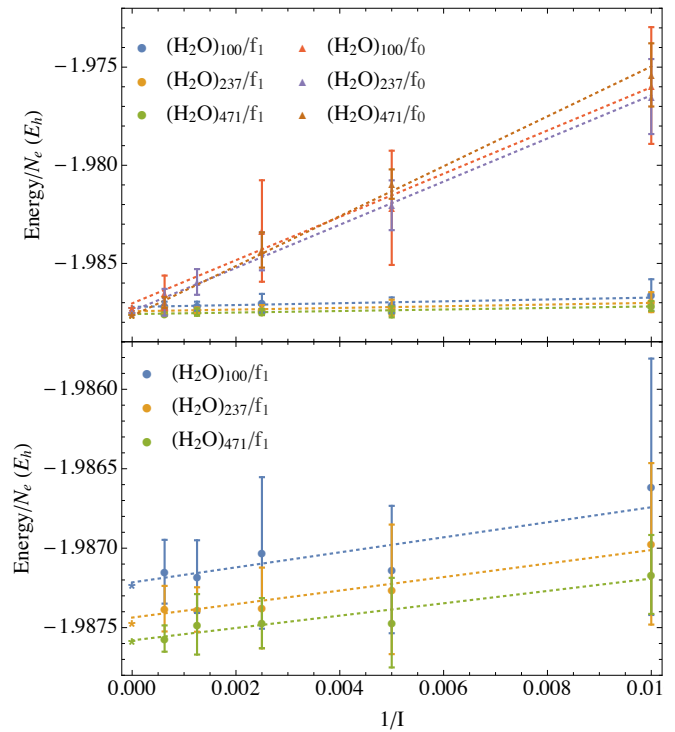}
\par\end{centering}
\caption{\label{fig:sDFTvalidation}Top panel: The estimate of energy per electron
as a function of the inverse number of random vectors ($1/I$) for
water molecule clusters of indicated sizes, without fragments ($/f_{0}$)
and with fragments (discussed in section~(\ref{sec:Embedded-fragments-based}))
of single $\text{H}_{2}\text{O}$ molecules ($/f_{1}$). The dotted
lines are linear fit to the data (weighted by the inverse error bar
length). The deterministic results are represented at $1/I=0$ by
star symbols. Bottom panel: a zoomed view of the $/f_{1}$ results.
These results were calculated using the STO-3G basis-set within the
LDA.}
\end{figure}

\begin{figure}
\begin{centering}
\includegraphics[width=1\columnwidth]{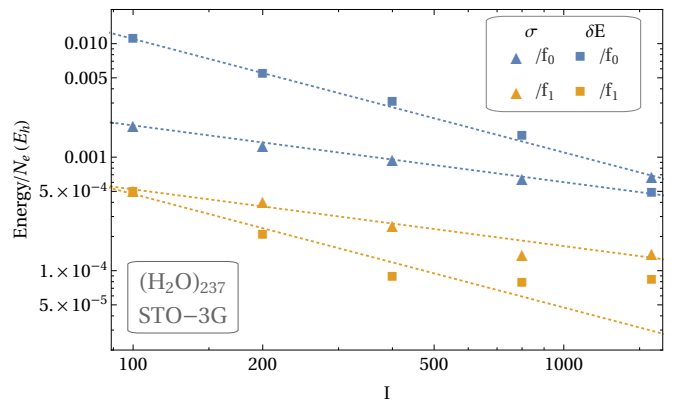}
\par\end{centering}
\caption{\label{fig:errors-vs-I}The standard deviation ($\sigma,$circles)
and errors ($\delta E$, squares) of the stochastic estimate of the
energy per electron as a function of the number of random vectors
($I$) in $\left(\text{H}_{2}\text{O}\right)_{237}$ without fragments
($/f_{0}$, blue) and with $\text{H}_{2}\text{O}$ fragments ($/f_{1}$,
yellow, discussed in section~(\ref{sec:Embedded-fragments-based}))
. The dashed lines are best fit functions $\alpha I^{-n}$to the data,
where $n=1/2$ for fitting the standard deviations and $n=1$ for
the bias. These results were calculated using the STO-3G basis-set
within the LDA.}
\end{figure}

Having demonstrated the validity of our deterministic numerical implementation
by comparing to deterministic DFT results of Q-CHEM, let us now turn
our attention to demonstrating the validity of the sDFT calculation
when comparing it to deterministic calculation under the same conditions.
In Fig.~\ref{fig:sDFTvalidation} (top panel) where we plot, for
water clusters of three indicated sizes, the energy per electron as
a function of $1/I$, where $I$ is the number of random vector $\chi's$
used for the stochastic trace formulas (Eq.~(\ref{eq:DensityAsStochTrace})-(\ref{eq:StochasticMoments})).
As the the number of random vectors $I$ grows (and $1/I$ drops)
the results converge to the deterministic values (shown in the figure
as stars at $1/I=0$). We repeated the calculations 10 times with
different random number generator seeds and used the scatter of results
for estimating the standard deviation $\sigma$ and the expected value
$\mu$ (these are represented, respectively, as error bars and their
midpoints in the figure). It is seen that the standard deviation in
the energy per particle drops as $I$ increases and in Fig.~\ref{fig:errors-vs-I}
it is demonstrated that the standard deviation drops as $I^{-1/2}$,
in accordance with the central limit theorem. The average values of
the energy per particle in Fig.~\ref{fig:sDFTvalidation} drop steadily
towards the converged deterministic values (stars). The fact that
the average is always larger than the exact energy, as opposed to
fluctuating around it, is a manifestation of a bias $\delta E$ in
the method. When $\delta E$ is larger than $\sigma$ it drops in
proportion to $I^{-1}$. In subsections (\ref{subsec:fluctuations})-(\ref{subsec:bias})
we will discuss and explain this behavior.


\subsubsection{\label{subsec:fluctuations}Statistical fluctuations}

\begin{figure}
\begin{centering}
\includegraphics[width=1\columnwidth]{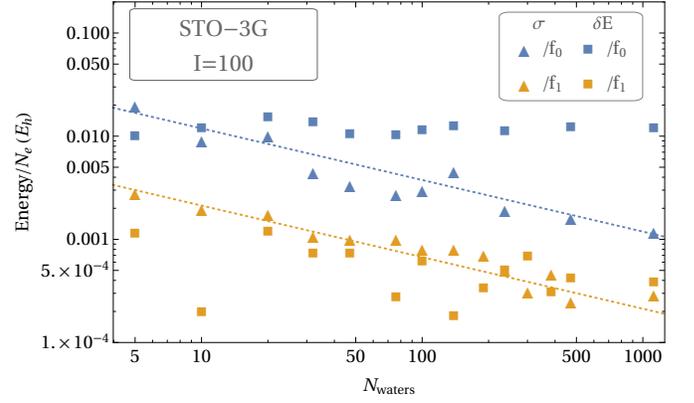}
\par\end{centering}
\caption{\label{fig:errors-vs-N}The standard deviation ($\sigma$, triangles)
and errors ($\delta E$ squares) of the stochastic estimate of the
energy per electron as a function of the number of water molecules
$N_{water}$ using no fragments ($/f_{0}$, blue markers) and single
water molecule fragments ($/f_{1}$, yellow markers). The dotted lines
are $1/\sqrt{N_{water}}$ fits to the $\sigma$ values. These results
were calculated using the STO-3G basis-set within the LDA and employed
$I=100$ random vectors. }
\end{figure}

It is straightforward to show that the variance of the trace formula
Eq.~(\ref{eq:HutchinsonTrace}) is: 
\begin{align}
\Sigma_{M}^{2} & \equiv\text{\textbf{Var}}\left\{ \sum_{\alpha\alpha'}^{K}\chi_{\alpha}M_{\alpha\alpha'}\chi_{\alpha'}\right\} \label{eq:HutchinsonVariance}\\
 & =\frac{1}{2}\sum_{\alpha\ne\alpha'}^{K}\left(M_{\alpha\alpha'}+M_{\alpha'\alpha}\right)^{2}\nonumber \\
 & =\left(\text{sym}\right)2\sum_{\alpha\ne\alpha'}^{K}M_{\alpha\alpha'}^{2},
\end{align}
where $\left(\text{sym}\right)$ marks an equality when $M$ is a
symmetric matrix. Therefore, from Eq.~(\ref{eq:DensityAsATrace})
the variance in the density $\hat{n}\left(\boldsymbol{r}\right)$\textbf{
}is 
\begin{align}
\text{\textbf{Var}}_{\text{I}}\left\{ \hat{n}\left(\boldsymbol{r}\right)\right\}  & =\frac{8}{I}\sum_{\alpha\ne\alpha'}^{K}\left[\sum_{\beta}^{K}P_{\alpha\beta}\phi_{\beta}\left(\boldsymbol{r}\right)\phi_{\alpha'}\left(\boldsymbol{r}\right)\right]^{2}.
\end{align}
The quantity inside the square brackets involves a limited number,
independent of system size, of $\alpha'$-$\beta$ index pairs $\left[\left(PS\right)_{\alpha\alpha'}\right]^{2}$
and since $tr\left[PS\right]=N_{e}$ we can assume that the magnitude
of the brackets squared is $O\left(\frac{N_{e}}{K}\right)^{2}$, i.e.
independent of system size. Summing over $\alpha$ introduces a system
size dependence, hence we conclude that $\text{\textbf{Var}}_{\text{I}}\left\{ \hat{n}\left(\boldsymbol{r}\right)\right\} $
has magnitude of $O\left(\frac{N_{e}}{I}\right)$. When the system
is large enough $P$ becomes sparse and then $\text{\textbf{Var}}_{\text{I}}\left(n\left(\boldsymbol{r}\right)\right)$
will tend to become of the magnitude $O\left(\frac{1}{I}\right)$,
i.e. system-size independent. The same kind of analysis applies to
any single electron observable $\hat{O}$ with sparse matrix representation:
\begin{equation}
\boldsymbol{\text{Var}}_{\text{I}}\left\{ \hat{O}\right\} \propto\frac{N_{e}}{I}.\label{eq:ExtensiveVariance}
\end{equation}
and independent of system size once $P$ localizes. Since intensive
properties are obtained by dividing the related extensive properties
by $N_{e}$, the standard deviation per electron of intensive properties
will evaluate as:
\begin{align}
\sigma_{\text{intensive}} & \propto\frac{\sqrt{\boldsymbol{\text{Var}}_{\text{I}}\left\{ \hat{n}\left(\boldsymbol{r}\right)\right\} }}{N_{e}}\propto\frac{1}{\sqrt{IN_{e}}}.\label{eq:IntensiveStDev}
\end{align}
The decay of the sDFT fluctuations with system size, first pointed
out in Ref.~\onlinecite{Baer2013}, is compatible with the fact that
fluctuations in intensive variables decay to zero in the thermodynamic
limit \citep{Gibbs1902}. For non-metallic systems $P$ becomes sparse
as system size grows. Once this sparsity kicks in, $\sigma_{\text{intensive}}$
is expected to decay as $1/\left(\sqrt{I}N_{e}\right)$. A numerical
demonstration of Eq.~(\ref{eq:IntensiveStDev}) is given in Fig.~\ref{fig:errors-vs-N},
for systems of varying numbers $N_{waters}$ of water molecules (all
using $I=100$ random vectors $\chi$), where the standard deviation
$\sigma$ in the energy per particle (blue triangles) indeed drops
with system size roughly as $N_{water}^{-1/2}$. 

\subsubsection{\label{subsec:bias}Bias due to nonlinearities}

In sDFT, the Hamiltonian $H=S^{-1}F$ is estimated using a random
density, and therefore it too, is a random variable with an expected
value $\bar{H}=\text{\textbf{E}}\left\{ H\right\} $ and a fluctuation
due to the a covariance matrix $\sigma_{ij;kl}^{2}=\boldsymbol{\text{E}}\left\{ H_{ij}H_{kl}\right\} -\bar{H}_{kl}$.
Consider an observable $\hat{O}$ with an exact expectation value
$\left\langle \hat{O}\right\rangle _{\bar{H}}=\text{Tr}\left(f\left(\bar{H};T,\mu\right)S^{-1}O\right)$
(Eqs.~(\ref{eq:P=00003Df(S^-1F)S^-1}) and (\ref{eq:<O>=00003D2E=00007Bchi^T*PO*chi=00007D})).
We note, that even when $\bar{H}$ is the exact (deterministic) Hamiltonian,
the expectation values $\left\langle \hat{O}\right\rangle _{H}$ will
\emph{not }average to the exact value $\left\langle \hat{O}\right\rangle _{\bar{H}}$,
simply because the function of the average of a random variable is
distinct or ``biased'' from the average of the function: $\text{\textbf{E}}\left\{ \left\langle \hat{O}\right\rangle _{H}\right\} \ne\left\langle \hat{O}\right\rangle _{\bar{H}}$.
Clearly, the extent of this bias stems from the how $\boldsymbol{\text{E}}\left\{ f\left(H;T,\mu\right)\right\} $
deviates from $f\left(\boldsymbol{\text{E}}\left\{ H\right\} ;T,\mu\right)$
and using Taylor's theorem this can be estimated as
\begin{align}
\text{\textbf{E}}\left\{ f\left(H;T,\mu\right)\right\}  & -f\left(\bar{H};T,\mu\right)=\label{eq:TheoreticalBias}\\
\frac{1}{2}\sum_{i,j,k,l} & \sigma_{ij;kl}^{2}\frac{\partial}{\partial H_{ij}}\frac{\partial}{\partial H_{kl}}f\left(\bar{H};T,\mu\right).\nonumber 
\end{align}
There are three lessons from this analysis: 1) all expectation values
$\left\langle \hat{O}\right\rangle $ based on $I$ random vectors
in the sDFT method suffer a \emph{bias} $\delta\left\langle \hat{O}\right\rangle \propto\text{coVar}_{\text{I}}\left\{ H\right\} \propto\text{Var}_{\text{I}}\left\{ \hat{O}\right\} $
; 2) from Eq.~(\ref{eq:ExtensiveVariance}) this bias in the \emph{intensive
}value $\left\langle \hat{O}\right\rangle /N_{e}$ is proportional
to $I^{-1}$ but independent from system size; and lastly: the double
derivative of $f$ on the right hand side of Eq.~(\ref{eq:TheoreticalBias})
(called the ``Hessian'') is related in a complicated way to the
curvature $f''\left(\varepsilon;T,\mu\right)$ of the Fermi-Dirac
function. This curvature is practically zero for almost all $\varepsilon$
except near $\varepsilon\approx\mu\pm k_{B}T$, and for sufficiently
small temperatures, the large Fermi-Dirac curvature regions are safely
tucked into the HOMO-LUMO gap, so that indeed the bias can be small.

Summarizing, we find the following trends in the SEs of intensive
quantities:
\begin{align}
\sigma_{\text{intensive}} & \propto\left(\frac{1}{N_{e}I}\right)^{1/2},\label{eq:stddev-vs-I,Ne}\\
\delta_{\text{intensive}} & \propto\frac{1}{I}.\label{eq:bias-vs-I,Ne}
\end{align}

Numerical demonstrations of Eq.~(\ref{eq:bias-vs-I,Ne}) are given
in Fig.~\ref{fig:errors-vs-I} (blue squares) where the bias $\delta E/N_{e}$
is seen to drop as $I^{-1}$ and in Fig.~\ref{fig:errors-vs-N} (blue
squares) where the bias is seen to be independent of the system size. 

\begin{figure*}
\includegraphics[width=1\columnwidth]{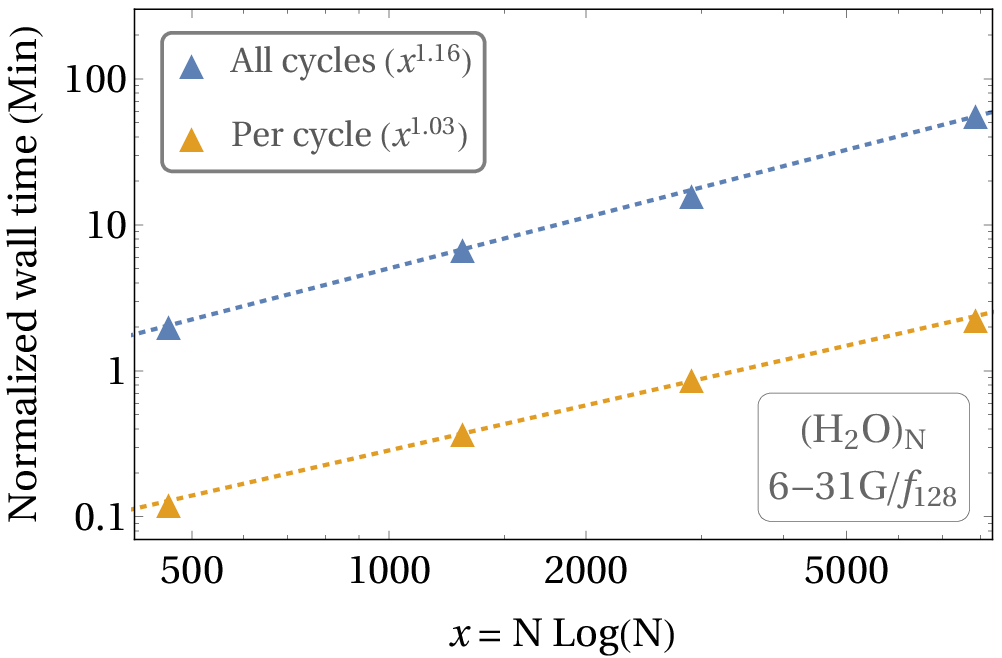}\includegraphics[width=1\columnwidth]{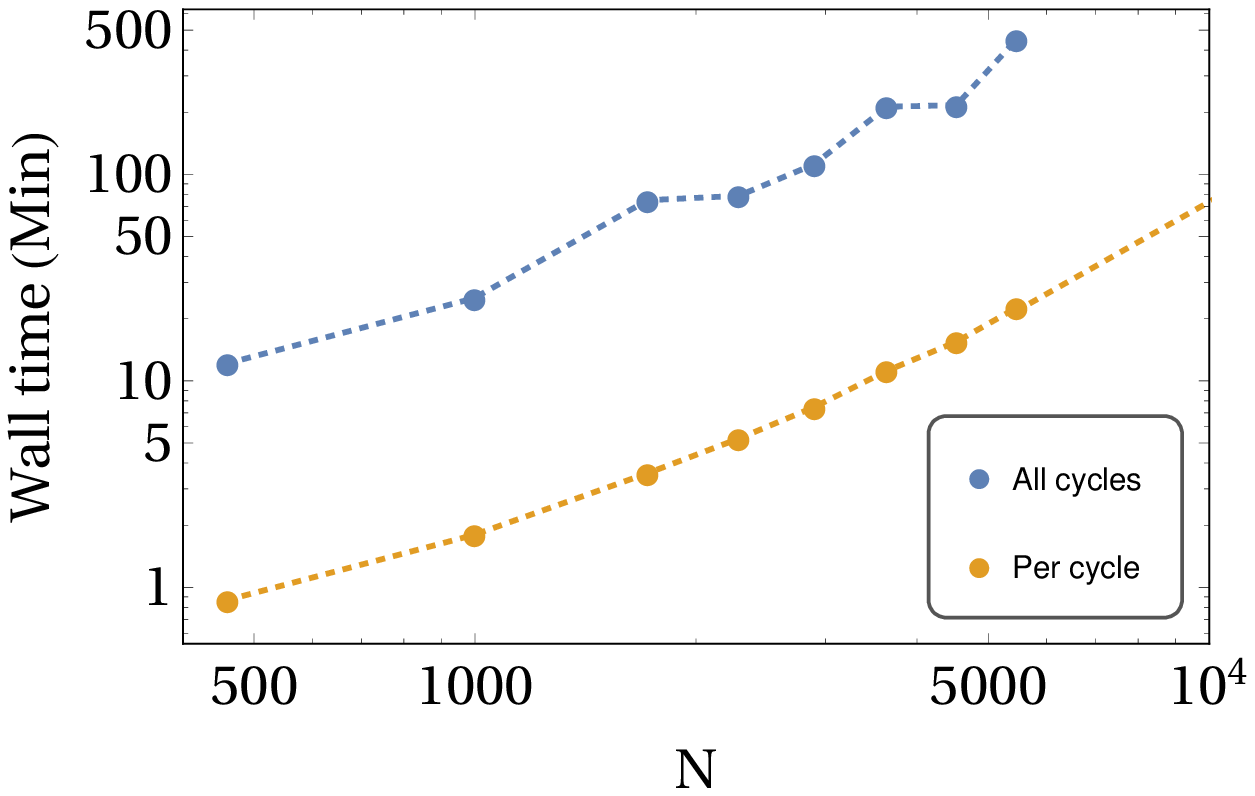}

\caption{\label{fig:scaling}The timing of DFT calculations of $\left(\text{H}_{2}\text{O}\right)_{N}$
water clusters using the 6-31G basis-set within the LDA. Left panel:
The sDFT wall time as function of $x=N\log N$ normalized to one random
orbital per thread for a full SCF calculation (blue symbols) and for
a single SCF cycle (orange symbols). Dashed lines are functions $t=Ax^{n}$,
where $n$ is best-fitted to the data and shown in the legend. Right
panel: Wall time of a conventional SCF calculation (using Q-CHEM \citep{Shao2015}),
performed on a single node, as a function of $N$ for a full SCF calculation
(blue symbols) and for a single SCF cycle (orange symbols). The Calculations
were run on an Intel Xeon CPU E3-1230 v5 @ 3.40GHz 64 GB RAM (\emph{without
}Infiniband networking). Each processor supports 8 threads. The sDFT
results were calculated with 800 random vectors and fragments of a
representative size of 128 water molecules (denoted $/f_{128}$). }
\end{figure*}

\subsubsection{\label{subsec:Scaling-and-scalability}Scaling and scalability }

In the left panel of Fig.~\ref{fig:scaling} we show, using a series
of water clusters how wall times scale as a function of system size
for the sDFT calculation. Because the evaluation of the Hartree potential
is made with fast Fourier transforms, the effort is expected to scale
as $x=N\log N$ where $N$ indicates the number of water molecules.
When considering a single SCF iteration we find this near-linear-scaling
as expected. When considering the entire calculation until SCF convergence
(which is achieved when the change in the total energy per electron
is smaller than $10^{-5}E_{h}$), we find the number of SCF iterations
growing gently with system size and the scaling seems to be near $O\left(x^{1.16}\right)$. 

\begin{figure}
\includegraphics[width=1\columnwidth]{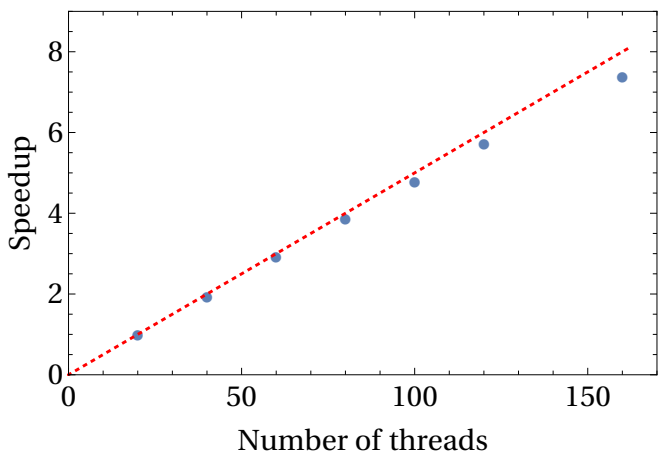}

\caption{\label{fig:scalability}Scalability plot of the calculation, showing
the speedup as a function of the number of threads used when calculating
a SCF iteration of $\left(\text{H}_{2}\text{O}\right)_{1120}$ (at
the 6-31G basis-set level within LDA) using a total of 2400 random
vectors. Calculations were performed on several 2.30GHz Intel Xeon
E5-2650 v3 with 252 GB and Infiniband networking. }
\end{figure}

As demonstrated in Fig.~\ref{fig:scalability}, we see excellent
scalability with number of processors with a  mere 8\% decrease from
the ideal speedup  when the number of cores was increased by a factor
of 8. This is a result of assigning to each thread a smaller number
of random vectors. Ideal wall times are achieved when the there is
but one stochastic orbital per thread \footnote{Once we apply one stochastic orbital per thread, further gain from
parallelization needs to be obtained from other sources, such as open
MP techniques. This has not been implemented yet.}. For the systems studied Fig.~\ref{fig:scaling} it is an hour for
a full SCF calculation of the water 1100 system (ca. 9000 electrons,
13000 orbitals). 

Under these conditions, the sDFT wall-times can be significantly lower
than those of ``conventional'' basis-set DFT calculations, as shown
in the right panel of Fig.~\ref{fig:scaling}. This happens despite
the fact that the program used, Q-CHEM, was remarkably still showing
quadratic scaling since the cubic scaling component was not yet dominant. 

\section{\label{sec:Embedded-fragments-based}Embedded fragments method}

\subsection{Theory}

The notion of fragments, developed first in Ref.~\onlinecite{Neuhauser2014a}
was to break up the system into disjoint pieces called fragments labeled
by the index $f$, and for each fragment compute a DM $P^{f}$, such
that to a good approximation we can write: 
\begin{equation}
P\approx\sum_{f}P^{f}.\label{eq:DM=00003DfDM}
\end{equation}
Clearly, the coherences between different fragments are also missing
from $\sum_{f}P^{f}$ and these too are assumed small but not totally
negligible. From Eq.~(\ref{eq:<O>=00003Dtr=00005BPO=00005D}), the
expectation value of an arbitrary one-electron operator $\hat{O}$
can be expressed as a contribution of two terms, $\left\langle \hat{O}\right\rangle =2\text{Tr}\left[\sum_{f}P^{f}O\right]+2\text{Tr}\left[\left(P-\sum_{f}P^{f}\right)O\right]$,
where the first is the ``fragment expected value'' and the second
is a \emph{correction}, expressed as a small trace to be evaluated
using the stochastic trace formula. Applying the stochastic trace
formula to just a small trace obviously lowers the SEs when compared
to using it for a full trace.

Ref.~\onlinecite{Neuhauser2014a} considered two types of fragmentation
procedures. The first was to used natural fragments which could just
be considered separately, for example, a single water molecule in
a water cluster or a single $\text{C}_{60}$ molecule in a cluster
of $\text{C}_{60}$'s. Since the molecules are not covalently bonded
they are weakly interacting and Eq.~(\ref{eq:DM=00003DfDM}) is expected
to be satisfied to a good degree (however, adjacent water molecules
can interact via hydrogen bonds and this may reduce the efficacy of
the single-molecule fragments, as discussed below). 

\begin{figure}
\includegraphics[width=1\columnwidth]{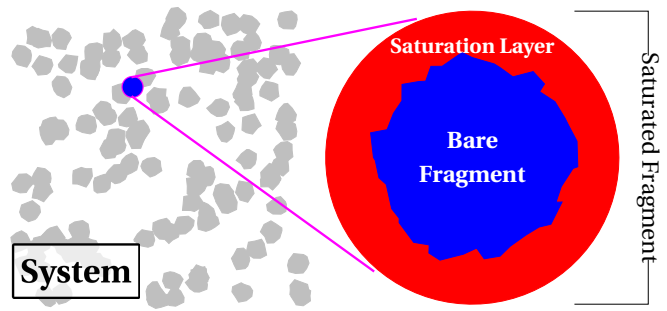}

\caption{\label{fig:Schematic-Fragment}A schematic depiction of a bare fragment
(blue region) as a localized set of atoms or molecules within the
large system. The fragment is first saturated by coating it with capping
atoms (red region), its \emph{saturated}-DM is calculated using a
deterministic DFT calculation, from which a bare DM $P^{f}$ is is
``carved'' out by an algebraic procedure. }
\end{figure}
The efficiency of the fragments depends entirely on the closeness
of the approximation in Eq.~(\ref{eq:DM=00003DfDM}) and therefore
significant effort has to go to developing techniques for constructing
fragments. One can probably make good use of the experience gained
by the biological and materials embedding methods \citep{Barone2010,Kamp2013,Sabin2010,Huang2011,Lan2015,Ghosh2010,elliott2010partition}.

The notion of \emph{saturated }fragments was developed further in
Ref.~\onlinecite{arnon2017equilibrium} and used in silicon clusters
where covalent bonds were cut when forming the \emph{bare fragment}.\emph{
}The \emph{dangling bonds }on the surface of the bare fragments were
then saturated with foreign H or Si atoms. This produced a saturated
fragment (see Fig.~\ref{fig:Schematic-Fragment}) and a special algebraic
technique was developed for carving out the bare fragment DM $P^{f}$.
The results facilitated what seems to be nearly unbiased force evaluations
for the atoms in large nanocrystals, with the structure studied using
Langevin molecular dynamics.

\begin{figure*}
\begin{centering}
\includegraphics[width=0.5\textwidth]{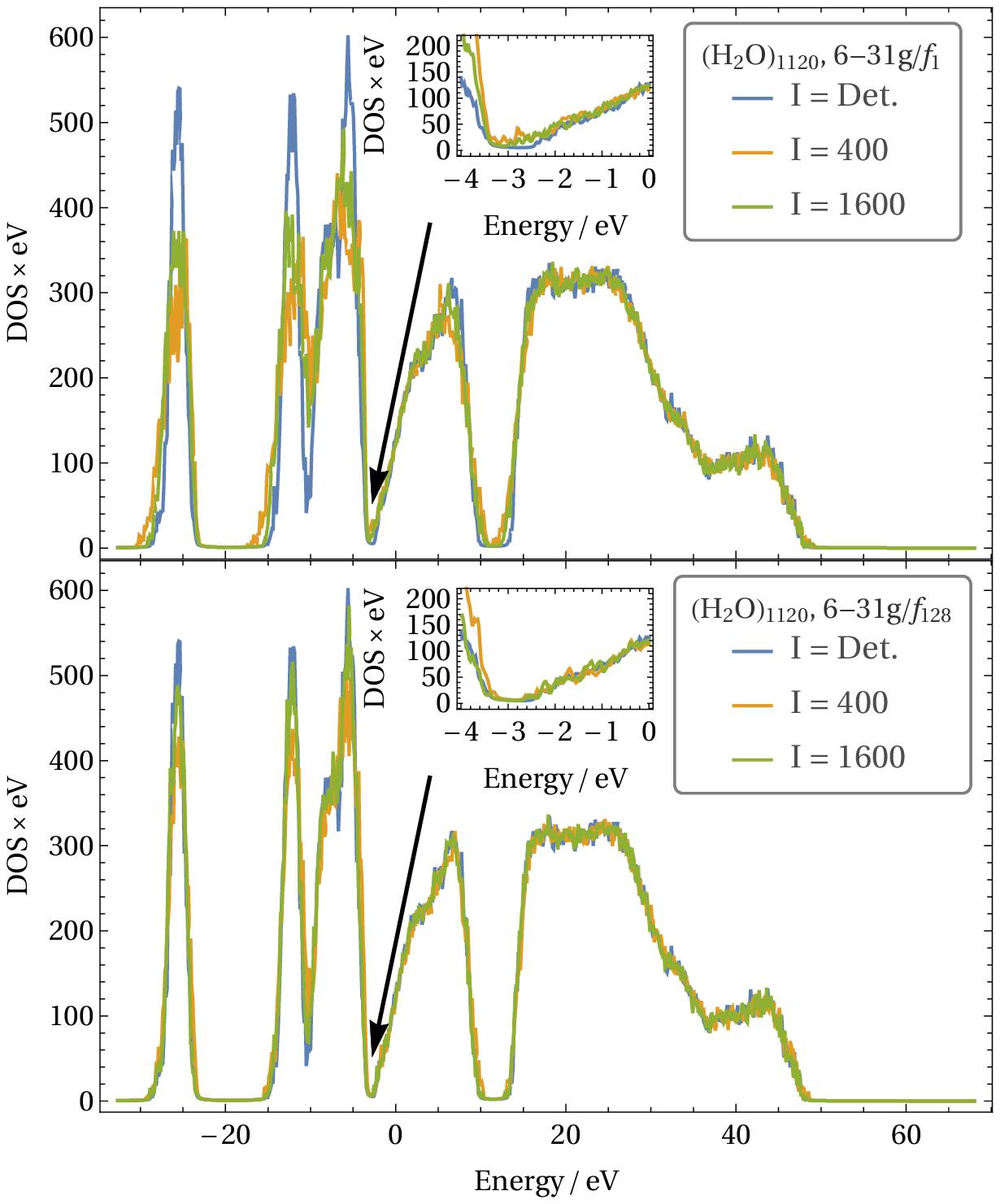}\includegraphics[width=0.5\textwidth]{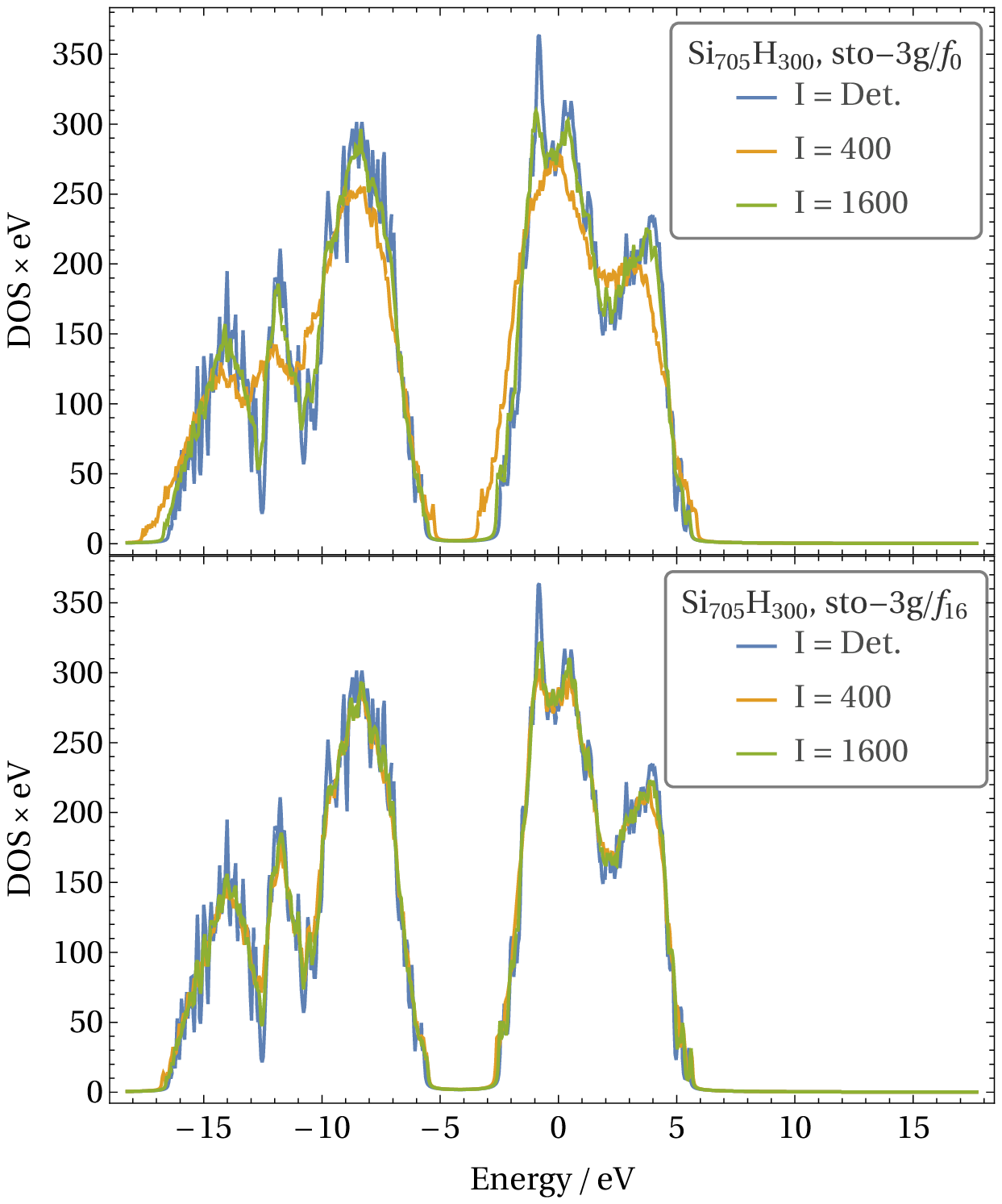}
\par\end{centering}
\caption{\label{fig:DOS}Left panels: The LDA DOS of a cluster of 1120 water
molecules using the 6-31G basis-set computed with $I=400$ and $I=1600$
random vectors and using single-molecule fragments ($/f_{1}$ top
panel) and 128 molecule fragments ($/f_{128}$ bottom left panel).
The insets zoom on the region of the band gap. Right panels: The LDA
DOS of $\text{Si}_{705}\text{H}_{300}$, a hydrogen-terminated silicon
nanocrystal, using the STO-3G basis-set computed with $I=400$ and
$I=1600$ random vectors and using no fragments ($/f_{0}$ top panel)
and 16 atom fragments ($/f_{16}$ bottom panel). In all panels the
results are compared to deterministic calculations under the same
conditions. }
\end{figure*}

\subsection{Efficiency of the embedded fragments}

To asses the utility of fragments that do not strictly require saturation,
such as water fragments in water clusters, consider first Fig.~\ref{fig:sDFTvalidation},
where we compare the energy per particle of $\left(\text{H}_{2}\text{O}\right)_{n}$,
with $n=100$, $237$ and $471$, estimated using sDFT with no fragments
(denoted $/f_{0}$) and using fragments of just one water molecule
($/f_{1})$. It is seen that there is a dramatic decrease in the the
standard deviation and in the bias. In Fig.~\ref{fig:errors-vs-I}
we study in more detail $\left(\text{H}_{2}\text{O}\right)_{n}$,
finding that with no fragments we are in a bias dominated regime while
the use of fragments allows us to move to a regime controlled by fluctuations.
Evidently, in the latter case, the large fluctuations mask the linear
decrease of the bias with $1/I$, which was so clearly visible in
the former one. In Fig.~\ref{fig:errors-vs-N}, we study the SEs
as a function of system size $N$, comparing the calculations with
and without fragments. We see that while fragments help reducing SEs,
they do not change the fact that the bias is largely independent of
$N$. 

The use of fragments greatly benefits other types of sDFT observables.
Consider, for example, the density of states function $\rho_{e}\left(E\right)$
of water \citep{Neuhauser2014a}. In the left panels of Fig.~\ref{fig:DOS}
we plot the DOS for a $\left(\text{H}_{2}\text{O}\right)_{1120}$
cluster described using the 6-31G basis-set comparing to the deterministic
result under an identical setup. We see in the top left panel, that
by using $I=400$ random vectors and small single-molecule fragments
($/f_{1}$), the sDFT DOS generally follows that of the exact calculation
quite closely. However, a zoom into the frontier orbital gap shows,
that even though the stochastic-based calculation exhibits, as it
should, a very low DOS in the frontier gap region, there is clearly
room for further improvement, since the gap is not sufficiently-well
described. Increasing the number of random vectors used from $I=400$
to $I=1600$ improves the overall accuracy but increasing the fragment
size to 128 water molecules ($/f_{128}$) is even more advantageous,
as can be seen in the lower left panel of the figure. It is evident
from this description that it is crucial to develop methods that enable
better fragments (in the sense that the approximation in Eq.~(\ref{eq:DM=00003DfDM})
is as tight as possible). Despite the obvious utility of the fragments
for the water cluster systems, there is a need to reach quite large
fragments for high accuracy. Perhaps this is due to the fact that
we do not saturate the bare fragments with their neighboring molecules,
as first suggested in recent unpublished work \citep{Ming2018}. Future
work will test this hypothesis.

Finally we also show in Fig.~\ref{fig:DOS} (right panels) the effect
of fragments on the DOS of a large silicon cluster. Here, we must
use saturated fragments, as was done in Ref.~\onlinecite{arnon2017equilibrium}.
The density of states, compared to a deterministic calculation is
again greatly improved when fragments of size 16 silicon atoms are
used (bottom left panel).

\subsection{\label{subsec:The-prospect-of}Localized energy changes }

So far, we have dealt with two types of observables: intensive properties
(such as energy per electron) which is a highly averaged quantity,
and density of states which, due to the tall number of levels in large
systems can be smeared, i.e. locally average, with little loss of
essential accuracy. We now demonstrate the possibility of calculating
forces on a small atom or molecule within the large system, using
stochastic DFT. Previous works concerning this issue \citep{Baer2013,arnon2017equilibrium}
demonstrated that the Hellman-Feynman force $F^{a}=-\int n\left(\boldsymbol{r}\right)\frac{\partial}{\partial R^{a}}v_{eN}\left(\boldsymbol{r}\right)d\boldsymbol{r}+\sum_{a\ne a}\frac{Z_{a}Z_{a'}\left(\boldsymbol{R}^{a}-\boldsymbol{R}^{b}\right)}{\left|\boldsymbol{R}^{a}-\boldsymbol{R}^{b}\right|^{3}}$
involves a controlled variance and small bias.

Here we consider a related but different question, the possibility
of systematically reducing the bias in forces on nuclei within a localized
region of interest in a large system. This is useful when modeling
reactions in biomolecular systems, as often done using the QM/MM approach,
where quantum chemistry forces are used for simulating chemical reactions
and other electronic processes (charge transfer or excitation) while
force fields are used for the rest of the system \citep{Cho2005,Svozil2006,Lin2007,Sharir-Ivry2008,Sabin2010,Barone2010,Kamp2013}.

\begin{figure*}
\includegraphics[width=1\textwidth]{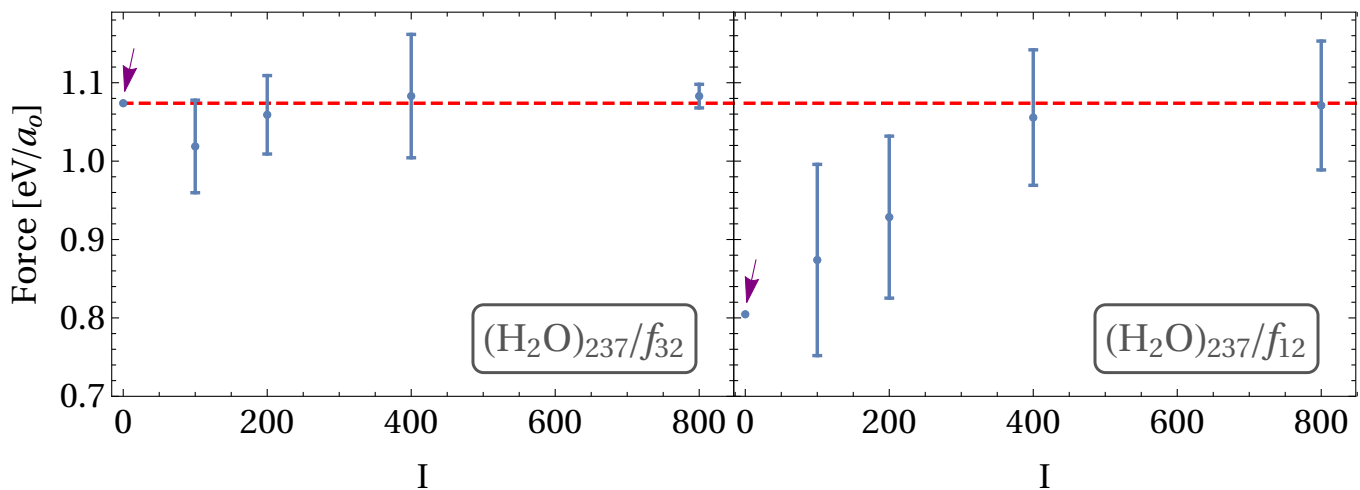}

\caption{\label{fig:WaterForce}The force on a marked water molecule in $\left(\text{H}_{2}\text{O}\right)_{237}$
(red dashed line is the deterministic DFT value) calculated as $-\delta E/\delta x$
where $\delta E$ is the energy difference between two positions of
the molecule displaced by a distance $\delta x=0.05a_{0}$. On the
left (right) panel we present $/f_{12}$ ($/f_{32}$) results. The
arrow points to the force $F\left(I=0\right)$ (the deterministic
force on the molecule when in the parent fragment). The ``error bars''
are 95\% confidence intervals for $\text{\textbf{E}}\left\{ F\left(I\right)\right\} $.
These results were calculated using the STO-3G basis-set within the
LDA.}
\end{figure*}
For this, we take a fragment which encapsulates the region of interest
and ``embed'' it into the system using sDFT. We study such a process
in Fig.~\ref{fig:WaterForce} where the force $F$ exerted on a certain,
marked, water molecule in a larger $\left(\text{H}_{2}\text{O}\right)_{237}$
cluster is calculated, first by deterministic DFT (shown as a dashed
red line in the figure) and then by sDFT as a function of the number
of random vectors $I$ ($F\left(I\right)$), using two types of fragment
sizes: 12-molecule fragments ($/f_{12}$), on the left, and larger
32-molecule fragments ($/f_{32})$ on the right. We note, that $F\left(I=0\right)$
is the deterministic force felt by the molecule in its parent fragment.
In the right panel we show the case of a parent fragment which fully
encloses the marked molecule. At $I=0$ the force is already very
close to the deterministic value, indicating $\sum_{f}P_{f}$ is an
excellent approximation for $P$. When embedded by $I>0$ stochastic
iterations, we find that fluctuations are introduced, but the error
bars (marking 95\% chance that $\text{\textbf{E}}\left\{ F\left(I\right)\right\} $
always include the exact value) indicate a small bias (such that the
error is not dominated by the bias). If we repeat this calculation,
but use small fragments which do not encapsulate the marked molecule,
the $F\left(I=0\right)$ is very different from the deterministic
exact force ($\sum_{f}P_{f}$ is a deficient approximation for $P$).
When embedded by $I>0$ stochastic iterations, the bias is gradually
removed as $I$ grows, in accordance with the steady diminishing of
the bias discussed in subsection \ref{subsec:bias}. 

We may conclude from this computational experiment that sDFT may be
especially useful for studying chemical processes in small subsystems
which can be encapsulated in fragments. Without using fragments, this
is also possible an increase in the number of samplings $I$ needs
to be employed in order to remove the bias.

\section{\label{sec:Summary}Summary and Discussion}

The sDFT approach has been used in various means and for a selection
of applications \citep{Baer2013,Neuhauser2014a,arnon2017equilibrium,Cytter2018,Neuhauser2015,Ming2018}.
The common thread for all the previous sDFT works was its formulation
using an orthogonal basis (grids or plane-waves representation). In
this review, we have focused on studying sDFT in the perspective of
a local non-orthogonal basis-set. One advantage of the localized basis-set
method is that even for large systems the deterministic calculation
can still be performed allowing to study in detail errors and their
dependence on system size. 

The sDFT theory was described using three stages, starting from the
standard basis-set formulation of DFT, leading to cubic scaling. Next,
we developed a deterministic trace-based calculation, exploiting the
sparsity of the Fock and overlap matrices, which lead to a quadratic
approach but remained numerically accurate. Finally, came the sDFT
which uses stochastic sampling to evaluate the trace-based calculations,
thereby lowering the scaling to linear. The price to pay is the introduction
of statistical errors, which one can mitigate by increasing the sampling
rate. In order to study and demonstrate the sDFT properties, we developed
a basis-set DFT approach using an auxiliary grid for constructing
the Hartree and exchange-correlation matrices. Based on this code
we also developed the stochastic sDFT implementation. We also developed
a basis-set-based fragment method and tested its utility

Using the code, we analyzed the statistical errors associated with
the stochastic calculations and their dependence on the number of
stochastic samples $I$, the system size, $N$ (one can take the number
of electrons $N_{e}$ or the basis-set size $K$ as $N$), and the
fragment size. As in previous sDFT papers, the results demonstrated
a $I^{-1/2}$ and $N^{-1/2}$ dependence of the statistical fluctuations.
Furthermore, we were able to explore the nature of the systematic
errors in the sDFT calculation. The bias errors in stochastic methods,
have been discussed before in \citep{Cytter2018,Neuhauser2017}. In
sDFT we show that they do not grow with system size and that they
decay as $I^{-1}$. We also developed an analytical model to explain
these observations. 

It has also been shown that using fragments the noise in the results
can be significantly reduced reaching a regime where the statistical
fluctuations are the dominating contributions to the error (rather
than the bias). These conclusions are in line with previous studies
\citep{Neuhauser2014a,arnon2017equilibrium,chen2018overlapped}, By
implementing the fragments we were able to calculate other observables
(such as the density of states, or forces) in a much more accurate
fashion for a very similar cost. 

We demonstrated that our sDFT implementation displays system-size
linear-scaling CPU time (Figure~\ref{fig:scaling}) and that it is
efficacious in parallel architectures (Figure~\ref{fig:scalability}).
Indeed, it seems to reach its full utility in CPU-abundant architectures,
suggesting it may be suitable for Exascale computing. 

Future work in the sDFT implementation is required for speeding up
the calculations on each node, this can be achieved by shared-memory
or GPU parallelization. Further development is also needed for improved
fragments which will reduce the variance and bias errors as well as
reduce the number of SCF iterations. Finally, as mentioned above,
using the sDFT code to drive a Langevin sampling of the nuclear configurations
\citep{arnon2017equilibrium} will allow us to compute observables
related to the thermal-nuclear structure of the molecular systems. 
\begin{acknowledgments}
R. Baer gratefully thanks Professor Yihan Shao of Univeristy of Oklahoma
for his continued support of our group's use of Q-CHEM. RB also acknowledges
ISF grant No. 189/14 for supporting this research. D. Neuhauser is
grateful for support by the NSF, grant DMR-1611382. E.R. acknowledges
support from the Physical Chemistry of Inorganic Nanostructures Program,
KC3103, Office of Basic Energy Sciences of the United States Department
of Energy under Contract No. DE-AC02-05CH11232.
\end{acknowledgments}

%

\end{document}